# Who teaches science in Alagoas? A quantitative analysis of non-specialist teaching at primary and secondary schools in Brazil

Robert Fischer[1], Elton Fireman[2], José Renan Gomes[2]

[1] *OPTMA, Universidade Federal de Alagoas, Maceió, Brazil*

[2] *CEDU, Universidade Federal de Alagoas, Maceió, Brazil*

Corresponding author: robert@optma.org, OPTMA, Caixa Postal 2051, CEP 57061-970, Maceió, AL, BRAZIL

Robert Fischer received his PhD in Physics at the Australian National University and then worked in industry. After managing a European project for the development of educational material, he moved to the Federal University in Alagoas and currently develops educational programs for the Federal Institute of Alagoas.

Elton Fireman received his PhD in Physics at the Universidade Federal de São Carlos. As a member of the Post-graduation Program for the Teaching of Science and Mathematics at the Federal University of Alagoas, he has been training science teachers in Alagoas for many years.

José Renan Gomes received a master degree in education at the Federal University of Alagoas, where he works as lecturer.

# Who teaches science in Alagoas? A quantitative analysis of non-specialist teaching at primary and secondary schools in Brazil

While non-specialist teaching increasingly becomes an object of public interest, there is little data available on the actual educational background of teachers giving classes outside their specialisation. This work analyses the data collected by the Brazilian Ministry of Education from public and private schools in Alagoas (Brazil), with a special focus on science education at primary and secondary schools. We find that the phenomena of non-specialist teaching is highly subject specific. For instance, while both maths and biology are little affected, more than half of physics classes are given by teachers without an education in any natural science.

**Introduction**

Non-specialist teaching (also known as out-of-field or foreign-subject teaching) at secondary schools appears to be a problem discussed more in public news-papers than in scientific literature. For example, Prof. Simon Margison from the Centre for the Study of Higher Education at the University of Melbourne writes in *The Australian* (20[th] of March 2013) that the country's future competitiveness is at risk because "one-third of our [Australian] secondary school classes in maths are taught by teachers untrained or undertrained in the discipline. Many senior science teachers are untrained or teaching the wrong science." In the US News (8[th] of August 2011), Jason Koebler discusses problems in science education at schools in the United States and cites the former assistant secretary at the U.S. Department of Education with the conclusion: "You can't teach what you don't know." Alom Shaha (The Guardian, 13[th] of December 2010) notes for the UK: "I fear I'll be meeting more and more people who will tell me they hated physics at school […] because, strictly speaking, these people won't ever have had an actual physics teacher." His argument is supported by Prof. Smithers, who is cited in the *Daily Mail* (17[th] of June 2012, by Laura Clark) saying: "If you have a biologist teaching physics, even at age 11, it may well be that their enthusiasm for physics isn't there, and the child isn't excited by it and moves in another direction". Julia Neuhauser points out (*Die Presse* ,1[st] of July 2012) that non-specialist has become common in Austria because the "Problembewusstsein" (awareness of it being a problem) is missing.

Such newspaper articles build on the limited information governmental bodies are willing to share with the public and often combine it with anecdotal evidence. With the notable exception of the "School Workforce Census" in the UK (DoE 2012) and the survey "SASS" in the US (Hill and Gruber 2011), there is little quantitative data available on the number and background knowledge of teachers who are teaching subjects foreign to their own specialization. One reason for the scarcity of data might be that the topics is "politically sensitive" (Ingersoll 2003): Those who are responsible for the managing and/or financing of schools, and hence in the position to provide this data, find it already difficult to employ a sufficient *number* of teachers to give the curses demanded by the curricula. A public discussion about the *subject-specific qualification* of educators would probably increase the pressure on these governmental bodies. Nonetheless, authors of public media articles – as e.g. those mentioned above – argue that content knowledge and personal interest in the teaching subject has an impact on the quality of education. They discuss especially the effect on so-called STEM subjects (<u>s</u>cience, <u>t</u>echnology, <u>e</u>ngineering, and <u>m</u>athematics), which many governmental and non-governmental institutions nowadays point to as particularly important for the development of their country. As cited above, some journalists even state that non-specialist teaching in STEM subjects eventually undermines the technological future and thus the welfare of their country. Unfortunately, up to now there is very limited quantitative and qualitative research to see if such concerns are justified, and how, if required, effective countermeasures should be designed.

Besides this public interest, the issue of non-specialist teaching deserves the attention of those providing teacher-training programs for an additional reason: Offering support to teachers who are interested in acquiring content knowledge on a foreign subject (which they might be requested to teach by their headmaster), may have a significant impact on the quality of education while requiring a relative low investment of time and effort. Considering that most of these teachers already possess didactic skills and classroom experience, the formation can focus on pedagogical content knowledge (Shulmann 1986) and help to apply it in the classroom (Ball 2000). Furthermore, such in-service support programs could offer quasi-immediate impact when compared to the training of new specialist teachers 'from scratch'.

In Brazil, this idea of a fast and easy 'fix' of imbalances in the teacher formation and teacher demand motivated the creation of the program "Segunda Licenciatura" (second graduation) in 2009 as part of the Plano Nacional de Formação de Professores da Educação Básica (PARFOR, in English: National plan of the teacher training for basic education). Within the framework of this "Segunda Licenciatura", teachers were offered a two-years full-time educational program to receive an additional degree for teaching a second subject. Since this program does not include any career related incentives (e.g. an increase in salary), the demand by teachers has been very low. Although other parts of PARFOR have been implemented in Alagoas by both private and public universities, no teacher has yet received (nor even been trained for) a degree in teaching an additional subject.

Our long-term objective is therefore the development of an improved approach to address the issue of non-specialist teachers in Alagoas. We believe that an increased acceptance and hence participation can be achieved if the program is tailored to the actual needs of non-specialist teacher, which in turn requires a better understanding of the difficulties non-specialist teachers are confronted with. Of particular interest in this context is the teacher's educational background. In order to identify relevant target groups, such as e.g. 'biology teachers teaching physics' or 'art teachers teaching chemistry', a precise quantitative analysis of the status quo for non-specialist teaching is required. The work reported in this article focuses on this particular point, relating the subjects taught with the areas the subject's teachers graduated in. To be more explicit, this paper does not claim that a mismatch in the teacher formation and the teacher employment has an impact on the quality of education, but aims at a deeper, quantitative, and differentiated understanding of the observed imbalances. A solid foundation of representative data is mandatory for future works studying the effects of non-specialist teaching on students, teachers and the qualitative outcome of the educational process.

The paper is structured as follows: After a brief introduction of the educational system in Brazil and the regional and historical context specific to the State of Alagoas, we will describe the origin and properties of the source data along with the methodology applied in the evaluation process. We then report the study results and discuss relevant findings and their implications on further research.

**Regional and historical context**

In contrast to Europe or the United States, the Brazilian primary and secondary school system open for the general public is very young. Although educational structures have been in place since the 18th century, education above alphabetisation was only affordable for a minority of the population. An educational system for the broad public was only build after a series of reforms and constitutional laws in 1961, 1971 and 1996. The most recent of these reform (1996) included a new law, which ensured the financial means to really expand the public school system and make free education equally available in all of Brazil. This in theory obligatory education – known as 'ensino fundamental' – encompasses primary and lower secondary education. The new law also regulated the so-called 'ensino médio', which corresponds to the upper secondary level in other countries, and aims at preparing students for entering university or the job market. Since 2001, conditional cash transfer programs (Bolsa Escola, Bolsa Família) have successfully encouraged poor families to send their children to school (Bruns, Evans, and Luque 2012).

Along with a notable improvement in the Brazilian educational system – reflected e.g. in improved ratings for Brazil's pupils in international comparative studies such as PISA (Bruns, Evans, and Luque 2012) – the rapid expansion of the public school system lead to a shortage of qualified teachers. In response to this shortage, a 'shortened' teacher training (licenciatura curta) was introduced in the 1970s (Prado and Hamburger 2001), which allowed teachers after only 2 years of university courses to graduate as teacher for the 'ensino fundamental', while 4 years were required to graduate as teacher for the 'ensino médio'. Today, teacher training for both level requires again a minimum of 4 years of training.

The Federal University of Alagoas started training physics, chemistry and biology teachers in 1974. While training for chemistry and biology teachers was offered also at state-run and public universities, physics teachers training courses were only offered at the Federal University. Significantly, in the 33 years between 1974 and 2007 only 65 students completed this course successfully (Barbosa 2006). Against this historical background, we expected to find a lack of physics specialist teachers, which was confirmed by the data presented below.

While maths and natural science subjects were traditionally a core part of the curriculum for upper secondary schooling in Brazil, the aspects of technology and engineering are taught at special federal institutions (Instituto Federal, previously known as Centro Federal Tecnológico), which offer upper secondary education in combination with professional training. However, the respective institution in Alagoas (IFAL) - although growing rapidly in the last few years – teaches only a rather small share of the overall ensino médio students in the state (see also Table 1).

Private schools, on the other hand, are common in Brazil and are generally regarded as offering a better education than free public schools. Since students from privately run schools often achieved significantly better results in the entrance exams of public universities than students from public schools, many parents regard tuition fees as a necessary investment even in face of free public alternatives.

**Source data and methodology**

Headmasters in Brazilian schools are required to register for every subject and grade taught at their school the teachers giving these classes. This data is collected annually by the Ministry of Education (MEC), and encompasses, among others, the teachers level and field of graduation. It has to be noted, though, that some teachers – mainly at private schools – are already teaching while still studying themselves for a university degree. For the particular focus of this study, namely "non-specialist teaching", these "studying teachers" were counted as subject specialists in the respective area they are currently studying. This decision was based on the assumption that such "studying teachers" show subject related interest and possess recent, though maybe incomplete, content knowledge in the field.

Since a subject-grade registration does neither include the total hours nor the number of students taught by a teacher (one registration unit may stay for a teacher teaching the subject to a small-size class or two or more, large classes at the same grade), the data is a somewhat indirect measure for the absolute quantity of non-specialist teaching. However, since students learn all subjects together with the same class members (students do not re-group for specific courses), this issue is expected to have a negligible impact on a comparison between subjects. On the other hand, the law-regulated form of data collection, large number of subject grade registrations and independent evaluation answer the concerns raised by Ingersoll (Ingersoll 2003) regarding studies measuring out-of-field teaching.

This work is based on the data from 2011 and encompasses a total of 30 463 registrations units recorded from public and private schools in Alagoas. More precisely, due to regulations in the data collection process, the data portrays the exact situation at Alagoas' schools at the 25$^{th}$ of March 2011. A total of 9 398 of these registrations refer to the level "ensino médio", which corresponds to the secondary level and were of primary interest for our study. We also studied further 2 937 units referring to "Ciências" (science) courses given at the state's private and public primary schools

("ensino fundamental") in the same year. The data distinguishes 4 categories of schools by their respective funding and managing organizations, namely private, municipal, state and federal schools (see Table 1).

|  | **Privada** (private Schools) | **Municipal** (Schools run by the local community) | **Estadual** (Schools run by the State of Alagoas) | **Federal** (Schools run by the Brazilian Government) | *Total* |
|---|---|---|---|---|---|
| **Ensino Fundamental** | 2718 | 11108 | 7239 | 0 | *21065* |
| **Ensino médio** | 2331 | 88 | 6617 | 362 | *9398* |
| *Total* | *5049* | *11196* | *13856* | *362* | **30463** |

Table 1: Source data. Number of registrations reported in 2011 at schools in Alagoas by educational level and funding/managing organization

**Results and Discussion**

As mentioned above, science is taught at Brazil's primary schools as an interdisciplinary subject, while natural science subjects are differentiated as physics, chemistry and biology only at "ensino médio". In order to avoid distortions in the data due to the particularities of a school system and deviating curricula (see Regional a Historical Context), we first concentrate on the registrations made by schools in the categories "Privada" and "Estadual" for " ensino médio", which together account for 95% of all registrations made for upper secondary schooling in Alagoas. We then analyze the educational background of teachers giving science classes at primary schools.

*Non-specialist teaching at upper secondary schools*

To create an overview of (non-)specialist teaching, Table 2 shows a matrix ordering the absolute number of registrations by subject and teacher specialization. A teacher's specialization (named "Teacher's Field of Graduation" in Table 2) is defined as the area in which the teacher has her highest degree or is currently pursuing a degree. Both subjects and areas of teacher specializations are arranged such that specialist teaching can be found roughly along the matrix diagonal. As a visual guide, green background indicates the highest values per subject while red frames mark subject specialist teaching.

| Teacher's Field of Graduation | Subject Taught by Teacher ||||||||||||||| total registrations per field of teacher specialisation |
|---|---|---|---|---|---|---|---|---|---|---|---|---|---|---|---|---|
| | IT | Maths | Physics | Chemistry | Biology | Geography | History | Philosophy | Social Sciences | Physical Education | Portuguese | English | Spanish | Other Foreign Language | Arts | Other | |
| Architecture | | 2 | | | | | | | | | | | | | 3 | | 5 |
| Engineering | | 19 | 56 | 11 | | | | | | | | | | | | | 86 |
| IT | 6 | | 3 | | | | | | | | | | | | 3 | | 12 |
| Maths | | 541 | 143 | 62 | 14 | | 2 | 4 | 6 | | | | 3 | | 10 | | 785 |
| Sciences | | 29 | 26 | 39 | 87 | 3 | | | | | 3 | | | | 3 | | 190 |
| Astronomy | | | 6 | 6 | | | | | | | | | | | | | 12 |
| Physics | | 10 | 132 | 10 | | | | | 1 | | | | | | 1 | | 154 |
| Chemistry | | 31 | 47 | 387 | 19 | 2 | 1 | 1 | 3 | 2 | 4 | 1 | 4 | | 7 | | 509 |
| Biology | | 32 | 30 | 78 | 604 | | 3 | | 3 | 12 | 2 | 6 | 1 | | 11 | | 782 |
| Agricultural Sciences | | 4 | 16 | 8 | 4 | | 3 | 3 | | | | 3 | | | | | 41 |
| Geography | 1 | 2 | 5 | 8 | 1 | 584 | 60 | 30 | 35 | 4 | 9 | 2 | 1 | | 20 | 1 | 763 |
| History | | 1 | 4 | 1 | 2 | 115 | 648 | 121 | 97 | 1 | 9 | 5 | 4 | | 53 | 1 | 1062 |
| Law | | | | | | 6 | 6 | 1 | 1 | | 7 | 5 | | | | | 26 |
| Philosophy | | | | | 1 | 6 | 8 | 98 | 46 | | 4 | 3 | 6 | | 7 | | 179 |
| Theology | | 3 | | | 1 | 7 | 50 | 25 | 2 | 2 | 1 | 10 | | | 17 | | 118 |
| Social Sciences | | | 1 | 1 | 4 | 82 | 54 | 22 | 39 | 1 | 7 | 1 | | | 5 | | 217 |
| Psychology | | | | | | | 1 | 29 | 30 | | | 15 | | | 3 | 1 | 79 |
| Physical education | | | 4 | | 2 | 5 | 6 | 1 | 1 | 496 | 1 | | | | 3 | | 519 |
| Portuguese | | 10 | 6 | 7 | 2 | 6 | 11 | 29 | 20 | 3 | 684 | 241 | 46 | 5 | 121 | | 1191 |
| Foreign Language | | 4 | 5 | 3 | | 7 | 2 | 14 | 17 | 3 | 315 | 491 | 125 | 2 | 105 | 1 | 1094 |
| Pedagogy | | 27 | 20 | 12 | 9 | 38 | 26 | 153 | 175 | 16 | 53 | 20 | 12 | | 64 | 3 | 628 |
| Arts | | | | | | | 6 | 3 | 3 | 4 | 6 | 6 | | | 177 | | 205 |
| Music | | | | | | | | | | | | | | | 15 | | 15 |
| Administration | | | 3 | | 4 | | | | | | | | | | | | 7 |
| Accounting / Economics | | 4 | 4 | | | | | | | | | | | | 1 | | 9 |
| Medicine | | 2 | 3 | 3 | | | | | | | | | | | | | 8 |
| Other | | 15 | 27 | 34 | 30 | 8 | 18 | 18 | 20 | 11 | 26 | 27 | 1 | 3 | 11 | 3 | 252 |
| total registrations per subject | 7 | 736 | 541 | 670 | 783 | 863 | 859 | 577 | 525 | 555 | 1132 | 827 | 213 | 10 | 640 | 10 | 8948 |

Table 2: Absolut numbers of registrations from Alagoas' private and state schools for 'ensino médio' in 2011, listed by subject and teacher specialization (area of teacher's highest graduation degree). Color-coding: as visual guide, high numbers per subject are marked with a green background, and subject-specialist teaching are marked with a red frame.

This matrix already allows a number of interesting observations, as for instance that physics is taught more often by maths teachers than by physicists, that social scientists rather teach geography than their 'own' subject, and that pedagogues are primarily employed to teach social sciences and philosophy. However, the variations in the total number of subject registrations make direct comparisons difficult.

A clearer picture is given when looking at the ratio of subject registrations for teachers with a degree in the specific subject (marked with red frame in Table 2) to the total amount of registrations for the subject (bottom row in Table 2). This specialist teaching rate has been calculated separately for public and private schools and is listed in percent in Table 3. Due to the low number of subject registrations, "IT" has been omitted, while the foreign languages have been combined to a single subject.

| | Maths | Physics | Chemistry | Biology | Geography | History | Philosophy | Social Sciences | Physical Education | Portuguese | Foreign Language | Arts | *Average* |
|---|---|---|---|---|---|---|---|---|---|---|---|---|---|
| Private Schools | 78.6 | 33.7 | 61.2 | 76.1 | 72.0 | 73.7 | 23.8 | 5.3 | 86.2 | 63.5 | 58.1 | 32.6 | *55.4* |
| Public Schools (Estadual) | 71.7 | 20.1 | 56.3 | 77.5 | 66.5 | 75.9 | 14.7 | 7.2 | 90.4 | 59.3 | 59.2 | 29.3 | *52.3* |

Table 3: Specialist teaching rate in percent for private and public secondary schools in Alagoas by subject.

While the pattern for non-specialist teaching is surprisingly similar, the data shows that privately run schools in general employ more specialist teachers than their public counterparts. It also shows that non-specialist teaching is a subject specific problem. Whereas both maths (73.5 %; this and the following rates are calculated for public and private schools combined) and biology (77.1 %) are taught predominately by specialists in the field, physics is affected particularly strongly. [Footnote: A comparison between physics and both social sciences and philosophy is misleading, since the latter two have been removed (e.g. during the period of the military dictatorship) and only been recently re-included into the national curriculum, hence discouraging prospective teachers to specialize in these fields.] In 2011, more mathematicians (26.4 %) were teaching physics than actual physicists (24.4 %). Chemistry, although having still a low share of specialist teachers (57.8%) when compared to subjects like history (75.4%), or even physical education (89.4%), lies in-between the two extremes cases of biology and physics.

The disparities between science subjects are clearly visible in Table 4, where we differentiate between non-specialist teaching of science subjects by scientists and non-scientists. For the purpose of this study "scientists" were arguably defined as teachers pocessing a degree in either physics, chemistry, biology, sciences or astronomy. For instance, the rate of non-scientists teaching biology is 10% or less at both private and public schools. The rate for chemistry is considerably higher. The data suggests that almost a quarter of chemistry lessons are given by non-scientists. However, more than half of the registrations for physics classes refer to teachers without a degree in any natural science. A good part of these non-scientists are mathematicians or engineers, but registrations also include language teachers and teachers holding a degree in accounting, administration or dentistry – specializations that apparently are only employed in maths and science teaching (see Table 2).

|  | Private Schools | | | | Public Schools (Estadual) | | | |
|---|---|---|---|---|---|---|---|---|
| **Taught by** | **Physics** | **Chemistry** | **Biology** | **Average** | **Physics** | **Chemistry** | **Biology** | **Average** |
| Subject Specialist | 33.7 | 61.2 | 76.1 | 57.0 | 20.1 | 56.3 | 77.5 | 51.3 |
| Other Scientist | 13.4 | 13.3 | 13.9 | 13.5 | 23.3 | 22.6 | 13.4 | 19.8 |
| Non-scientist | 52.9 | 25.5 | 10.0 | 29.5 | 56.6 | 21.1 | 9.1 | 28.9 |

Table 4: Specialist teaching rates in percent for natural science subjects.

Although at private schools the specialist teacher rate for physics is higher than at public schools, it has to be noted that the educational level of physics teachers at private schools is the lowest of all subjects: Only 77.3 % of those teaching physics at private schools possess or pursue a tertiary degree, while the average for all subjects lies at 90.3% for private and 96.9% at public schools.

One would expect that the specialist-teacher rate correlates with the availability of teachers for a specific subject. Since we have no data on unemployed or otherwise employed teachers in Alagoas, we estimate the relative availability of a subject specialist by dividing the total number of registrations for teachers with a degree in a given subject (right column in Table 2) by the total number of registrations for the same subject (bottom row in Table 2). A value of 1 would therefore indicate that all subject courses could be given by specialists, while a value higher or lower than 1 indicates an over or undersupply of subject specialists, respectively. To give an example: If there are only 154 registrations referring to teachers with a degree in physics but 541 registrations for the subject physics, it is to be expected that the specialist teaching rate for physics is low. In Figure 1 we plot this value of specialist availability against the actual specialist teaching rate. Two clusters can be observed, separated by a gap. One cluster, which includes physics, philosophy, arts, and social sciences, stands for 'problematic' subjects with low specialist availability and low specialists teaching rate, while the second one includes subjects with a high availability or even an oversupply of specialists. The diagonal line marks the theoretical "optimal" employment for a given availability of subject teachers. The closer a subject lies to the line, the more teachers who graduated in the subject also teach it. A good example is physical education, which is taught almost entirely by specialists in the field, who, in turn, teach little else than physical education. The graph suggests that a higher subject-specialist teaching rate was possible for maths, biology – both having an 'availability' ratio of close to 1 – and even chemistry (availability ratio: 0.76). Such a theoretical interpretation does of course not account for external factors, as for example the situation of a maths teacher in a small rural school who might rather choose to teach an additional subject at his school than maths at another, distant school lacking specialists. However, the graph makes clear that physicists are already focused on teaching "their" subject, so that an increase in the total number of teachers with a suitable physics content knowledge is needed to improve the subject specialist rate.

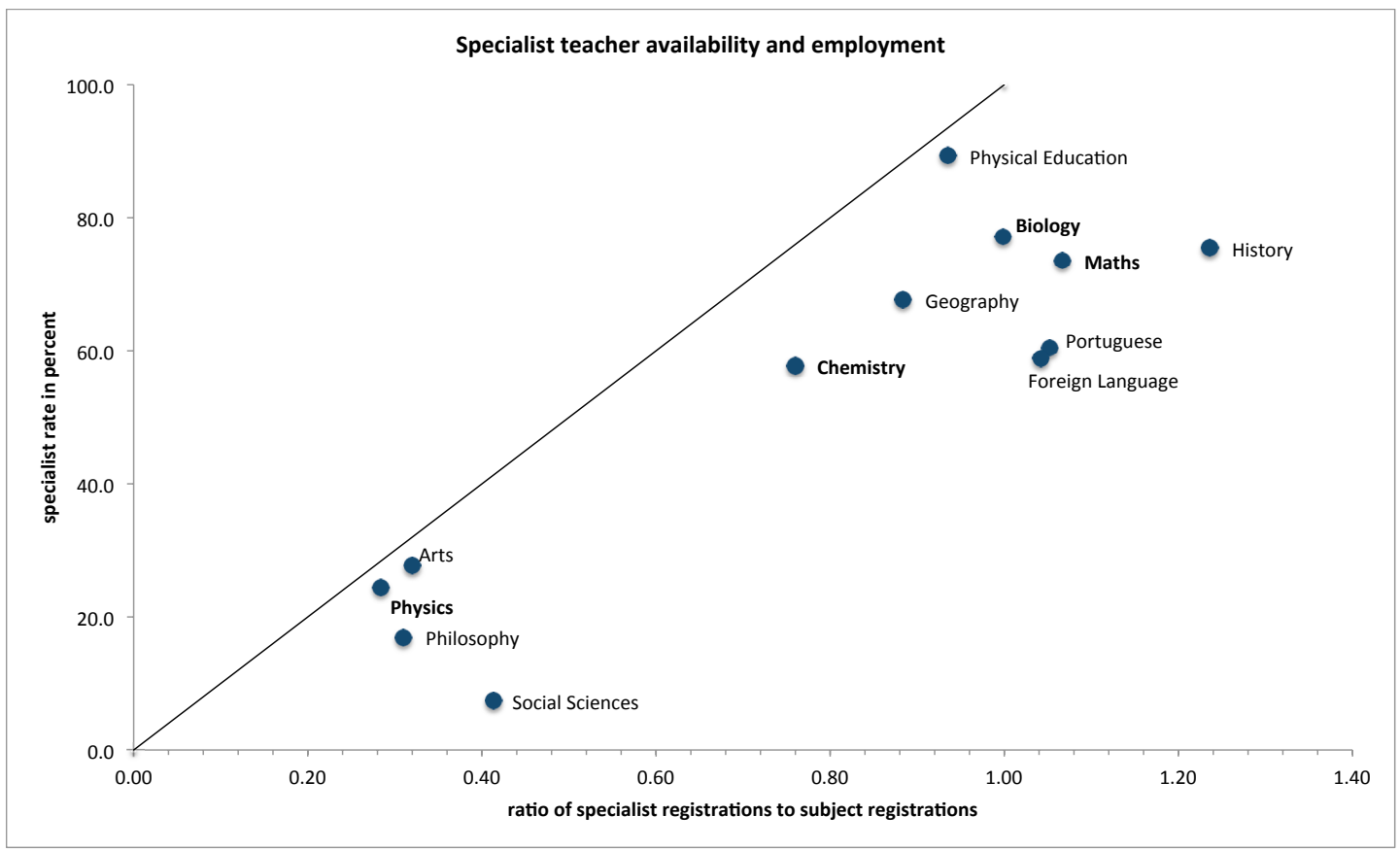

Figure 1: Subject specialist rate vs. the availability of subjects specialist teachers, estimated as the ratio of total specialist registrations to total subject registrations for each subject.

*Science teaching at primary schools*

Although Brazilian universities offer courses specifically to graduate as science teacher, the greater part of science classes at Alagoas primary schools are given by teachers with a different educational background. In fact, only 11.0 % of the registrations made for science classes indicate teachers who graduated in this subject. More than half of the registrations are made for non-scientists, yet with a lower share of mathematicians than found at secondary schools for science subjects. The scarcity of physicists can also be observed at "ensino fundamental": More geography (6.2 %), history (4.3 %) or language teachers (3.7 %) are teaching science than physicists (2.7 %). While chemists too are underrepresented (4.1 %), 30.0% of science classes are given by biologists (Figure 2). Obviously, this raises the question what impact such an imbalance of educational backgrounds of science teachers has on the impression primary school students get about science and STEM oriented careers. However, this question is beyond the scope of this work and will be studied in another research project.

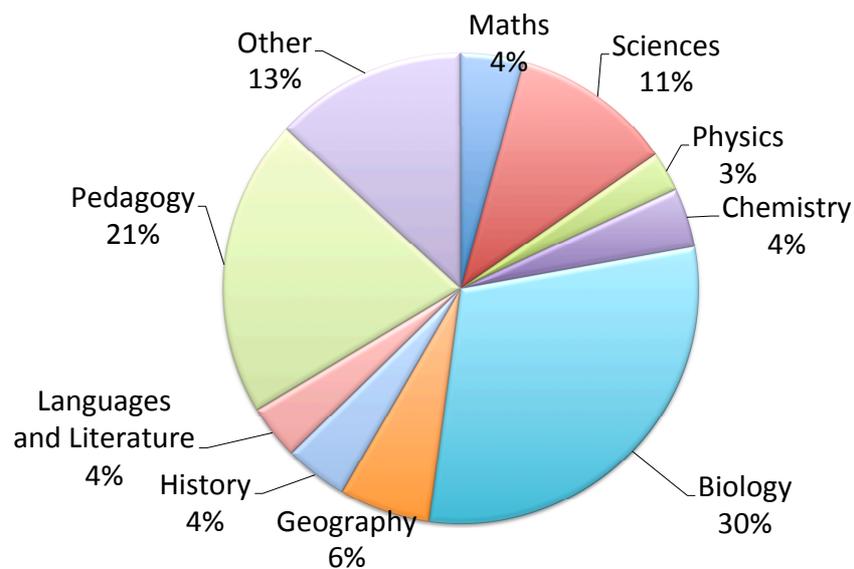

Figure 2: Area of graduation of educators teaching science at primary schools

**Conclusion**

In this work, we quantitatively analyzed the dimension of non-specialists science teaching at primary and secondary schools in Alagoas, Brazil. The data, collected in the year 2011, clearly shows that the problem is subject specific. While roughly ¾ of the biology and maths classes are taught by specialists in the field, chemistry and especially physics are often taught by non-specialists or even non-scientists. More than half of the "Ciências" (science) classes, a subject taught at Brazilian primary schools, are likewise taught by non-scientists.

This research has been motivated by our intention to develop tailored support programs for non-specialist teachers who are interested in improving content knowledge and subject specific didactics. Although the numbers reported above do not tell us what specific impact non-specialist teaching may have in a particular case, they do help us to identify some potential target groups. Since physics has been especially affected by non-specialist teaching, it deserves primary attention. The data reveals that non-specialists physics teachers at secondary schools often hold a degree in maths or engineering – a valuable knowledge base, which accordingly designed support programs could build on. Non-specialist teachers in chemistry, on the other hand, often graduated in a science subject, predominately in biology. Teachers giving 'science' classes at primary schools very often have no background in any STEM related field, and thus may require a far more comprehensive program to build up pedagogical content knowledge.

Although we briefly discussed the necessary historic and cultural background to provide the context for understanding the data, we do believe that the picture in other, higher developed school systems e.g. in Europe will show similar imbalances in teacher specialization and employment – notably in the area of natural sciences. Interestingly enough, the numbers reported by the UK Department of Education for the same year (2011) show that at public funded schools in the UK only 56.1% of physics teachers, 65.8 % of chemistry teachers and 76% of biology teachers hold a degree in the respective subject they teach (DoE 2012, Table 13). This seems to match the pattern seen in Table 3, although at a higher level.

We hope that the data provided in this work encourages colleagues to study the situation in other school systems, and, where found necessary, prepare initiatives to address the issue.

**References**

Ball, D. L.. 2000. *Bridging practices: Intertwining content and pedagogy in teaching and learning to teach.* Journal of Teacher Education, 51(2), pp. 241-247.

Barbosa, J.I., Serra, K., and Fireman, E.C.. 2006. *A formação do professor de Física na UFAL: as intenções e preocupações*,

Bruns, B., Evans, D. and Luque, J. 2012. *Achieving World-Class Education in Brazil, The Next Agenda*, published by the World Bank, DOI: 10.1596/978-0-8213-8854-9

DoE. 2012. *School workforce in England: November 2011*, published by the Department of Education, SFR 06/2012

Hill, J.G. and Gruber K.J.. 2011. *Education and Certification Qualifications of Departmentalized Public High School-Level Teachers of Core Subjects: Evidence From the 2007–08 Schools and Staffing Survey*, published by the US Department of Education, NCES 2011-317

Ingersoll R.M.. 2003. *Out-of-Field Teaching and the Limits of Teacher Policy*, research report published by the University of Washington, R-03-5

Prado, F. D., and Hamburger, E. W. 2001. *Estudos sobre o curso de Física da USP em São Paulo.* In: NARDI, R.(org.). Pesquisa em Ensino de Física. Série: Educação para Ciência. V. 1, 2ª edição revisada: Ed. Escrituras, 2001, São Paulo.

Shulman, L. S.. 1986. *Those who understand: Knowledge growth in teaching*. Educational Researcher, 15(2), 4- 31.